\documentclass[sigconf]{acmart} 
\AtBeginDocument{%
  }

\usepackage[toc,page]{appendix}

\usepackage{tikz}
\usepackage{amsmath}

\usepackage{filecontents}

\usepackage{amsmath,amsfonts}

\usepackage{amssymb}

\usepackage{enumitem}
\usepackage{algorithmic}
\usepackage{graphicx}
\usepackage{textcomp}
\usepackage{xcolor}
\usepackage{multirow}
\usepackage{makecell}
\usepackage{booktabs}
\usepackage{float}
\usepackage{epstopdf}
\usepackage[ruled,linesnumbered]{algorithm2e}
\usepackage{algorithmic}
\usepackage{color}
\usepackage{listings}
\usepackage{booktabs}
\usepackage{caption}
\usepackage{graphicx,subfig}
\usepackage{threeparttable}
\usepackage{amsmath}
\usepackage{pifont}
\usepackage{xcolor}
\usepackage{balance}
\definecolor{green}{HTML}{00A64F} 
\usepackage[utf8]{inputenc}
\usepackage{cleveref}
\usepackage{colortbl}

\setcopyright{acmlicensed} 
\copyrightyear{2026} 
\acmYear{2026} 
\acmDOI{XXXXXXX.XXXXXXX} 
\acmConference[Conference]{}{June 03--05, 2018}{Woodstock, NY}  
\acmISBN{978-1-4503-XXXX-X/2018/06, This manuscript reports preliminary findings.}  

\author{Liantao Song}
\affiliation{%
  \institution{National University of Defense Technology}
  \city{Changsha}
  \country{China}}
\email{songliantao@nudt.edu.cn}

\author{Yiming Zhang}
\affiliation{%
  \institution{Guangdong Polytechnic Normal University}
  \city{Guangzhou}
  \country{China}}
\email{yimingz@gpnu.edu.cn}

\author{Fengwei Zhang}
\authornote{corresponding authors.}
\affiliation{%
  \institution{Southern University of Science and Technology}
  \city{Shenzhen}
  \country{China}}
\email{zhangfw@sustech.edu.cn}

\author{Yan Ding}
\authornotemark[1]
\affiliation{%
  \institution{National University of Defense Technology}
  \city{Changsha}
  \country{China}}
\email{yanding@nudt.edu.cn}

\author{Bin Zhou}
\affiliation{%
  \institution{National University of Defense Technology}
  \city{Changsha}
  \country{China}}
\email{binzhou@nudt.edu.cn}

\author{Jie Yu}
\affiliation{%
  \institution{National University of Defense Technology}
  \city{Changsha}
  \country{China}}
\email{yj@nudt.edu.cn}

\author{Yusong Tan}
\affiliation{%
  \institution{National University of Defense Technology}
  \city{Changsha}
  \country{China}}
\email{ystan@nudt.edu.cn}

\begin{document}

\title{Shielded but Lightweight: Building Practical Confidential Containers with ARM CCA} 

\begin{abstract} 
 The rapid advancement of cloud-native technologies has created an urgent need for security. Currently, confidential containers are increasingly deployed in multi-tenant environments. Existing confidential container designs mainly adopt a microVM-based architecture. Although this approach improves inter-container isolation, its complex software stack leads to high startup latency and significant resource overhead, making it unsuitable for short-lived container workloads. In this paper, we propose Fasco, a lightweight confidential container runtime based on the ARM Confidential Compute Architecture (CCA). Fasco directly instantiates each container as an independent Container Realm, leveraging CCA's hardware-enforced isolation to ensure the confidentiality and integrity of application data inside the container. In addition, Fasco introduces a dedicated System Realm to provide system services and resource management for container realms. Through exception forwarding and shared buffers, Fasco ensures isolation among different container realms. We have implemented a prototype of Fasco and evaluated its performance on ARMv8 hardware. Experimental results show that Fasco reduces the startup latency and performance overhead of existing confidential container architectures while maintaining a small TCB.
\end{abstract}


\maketitle

\section{Introduction}
\label{1}

With the widespread adoption of cloud-native technologies, containers, by virtue of their ability to build and run elastic and scalable cloud applications, have become a key virtualization infrastructure at the foundation of cloud platforms. However, traditional containers rely on the host kernel to provide isolation mechanisms such as namespaces and cgroups, and therefore still essentially share the same operating system kernel. Once the host operating system or hypervisor is compromised, an attacker may directly read container memory, tamper with its execution state, or even exploit kernel vulnerabilities to achieve container escape. At present, confidential computing technologies are developing rapidly. By leveraging trusted execution environments (TEEs) provided by CPUs, they ensure the security of data and code during processing through hardware-based isolation and memory encryption. Typical TEEs include Intel SGX/TDX and AMD SEV. By building confidential containers on top of TEEs, it is possible to provide hardware-backed isolation protection for containerized workloads while preserving the lightweight nature of containers and compatibility with the existing ecosystem.

In recent years, ARM has introduced the Confidential Compute Architecture (CCA), a new type of TEE that supports the creation of Realm virtual machines—hardware-isolated execution environments that are protected against attacks from the host operating system and hypervisor. Existing confidential containers mainly adopt a design based on “encapsulating containers inside confidential virtual machines,” that is, packaging each container into a lightweight virtual machine (such as Coco) and then protecting it with confidential computing hardware such as AMD SEV, Intel TDX, or ARM CCA. However, this method also has obvious limitations. On the one hand, virtual-machine-based encapsulation inevitably introduces additional components such as a guest kernel, root filesystem, image management, and boot chain, causing confidential containers to incur significantly higher startup latency, memory usage, and storage overhead than ordinary containers. This makes them less suitable for typical cloud-native scenarios characterized by short container lifecycles, rapid creation and destruction, and elastic scaling. On the other hand, there is a mismatch between VM-level isolation and container-level abstraction in terms of isolation granularity, trust boundaries, and attestation semantics, which may lead to security issues.

In addition to the mainstream approach based on lightweight virtual machines, some studies have explored other ways to implement confidential containers. For example, RContainer and Shelter rely on trusted firmware or high-privilege monitors to directly intervene in the deployment and execution of containers, thereby enabling lower-level control over resource isolation and security policies. However, such methods increase the amount of trusted firmware code; once a high-privilege monitor contains vulnerabilities or is exploited, it may cause security damage on a much larger scale, resulting in a Trusted Computing Base (TCB) that is excessively large and fragile. Other approaches adopt unikernels, specialized runtimes, or highly customized application execution environments to reduce the size of the software stack and strengthen isolation. Yet these methods often suffer from clear limitations in POSIX compatibility, application migration cost, and ecosystem support, making it difficult for them to directly host existing cloud-native applications and OCI container images. Therefore, how to achieve a lightweight architecture design tailored to confidential-container granularity without significantly sacrificing compatibility remains an important issue in current confidential container research.

To address this, this paper proposes the Fasco design, a lightweight confidential container runtime for ARM CCA. It leverages the hardware-enforced isolation provided by Realms, allowing protected workloads to run independently of an untrusted host operating system and hypervisor. However, implementing practical confidential containers on ARM CCA faces two technical challenges: (i) minimizing the size of the TCB inside a Realm while still supporting existing applications in secure containers; and (ii) since containerized applications are generally lightweight and short-lived, enabling fast startup to reduce container deployment time.

To solve these problems, Fasco directly builds container-granularity confidential execution domains on ARM CCA, instantiating each container as an independent Container Realm in which it runs. It introduces a controlled System Realm responsible for resource coordination and mediation of system services. At the same time, it uses the RMM to provide memory isolation between confidential containers, and ensure the security of cross-domain interactions. Compared with existing microVM-based confidential container solutions, Fasco does not require attaching a full virtual-machine software stack to each container. As a result, while preserving the lightweight characteristics of containers and compatibility with OCI standards as much as possible, it can effectively reduce startup latency and resource consumption of confidential containers, making it better suited to cloud-native scenarios that demand containers to be lightweight, fast, and elastically scalable.

 Our main contributions are summarized as follows:
\begin{itemize}
    \item We propose a new lightweight confidential container architecture for ARM CCA. We present Fasco, a runtime architecture that builds confidential containers directly on ARM CCA. Unlike existing solutions that mainly rely on microVMs or the container-in-VM model, Fasco directly instantiates each container as an independent Container Realm and introduces a System Realm as a controlled management and service intermediary, thereby shifting the protection boundary of confidential computing from the VM level down to the container level.

    \item We design and implement a secure execution mechanism centered on the RMM. By extending the RMM, Fasco supports the creation of container Realms, image loading, page-table isolation, context switching, and exception interception. It also provides isolation and resource management mechanisms for multi-container scenarios. Through System-Realm-controlled exception forwarding, shared buffers, and result verification mechanisms, Fasco enables secure access to system services such as files, networking, and I/O.

    \item We implement a prototype on a real ARM platform and demonstrate its practicality and performance advantages. We implemented a Fasco prototype and deployed and evaluated it in both a QEMU CCA environment and on an RK3588 hardware platform. Experimental results show that, while preserving container-level confidentiality and integrity protection, Fasco delivers strong performance in system benchmarks, web services, and multi-container concurrent scenarios compared with existing microVM-based confidential container solutions, and significantly reduces startup latency and runtime overhead.

\end{itemize}

\section{Preliminaries and Motivation}
\label{2}

\subsection{Existing Container Architectures}

\begin{figure}[t]
    \centering    \includegraphics[width=0.95\linewidth]{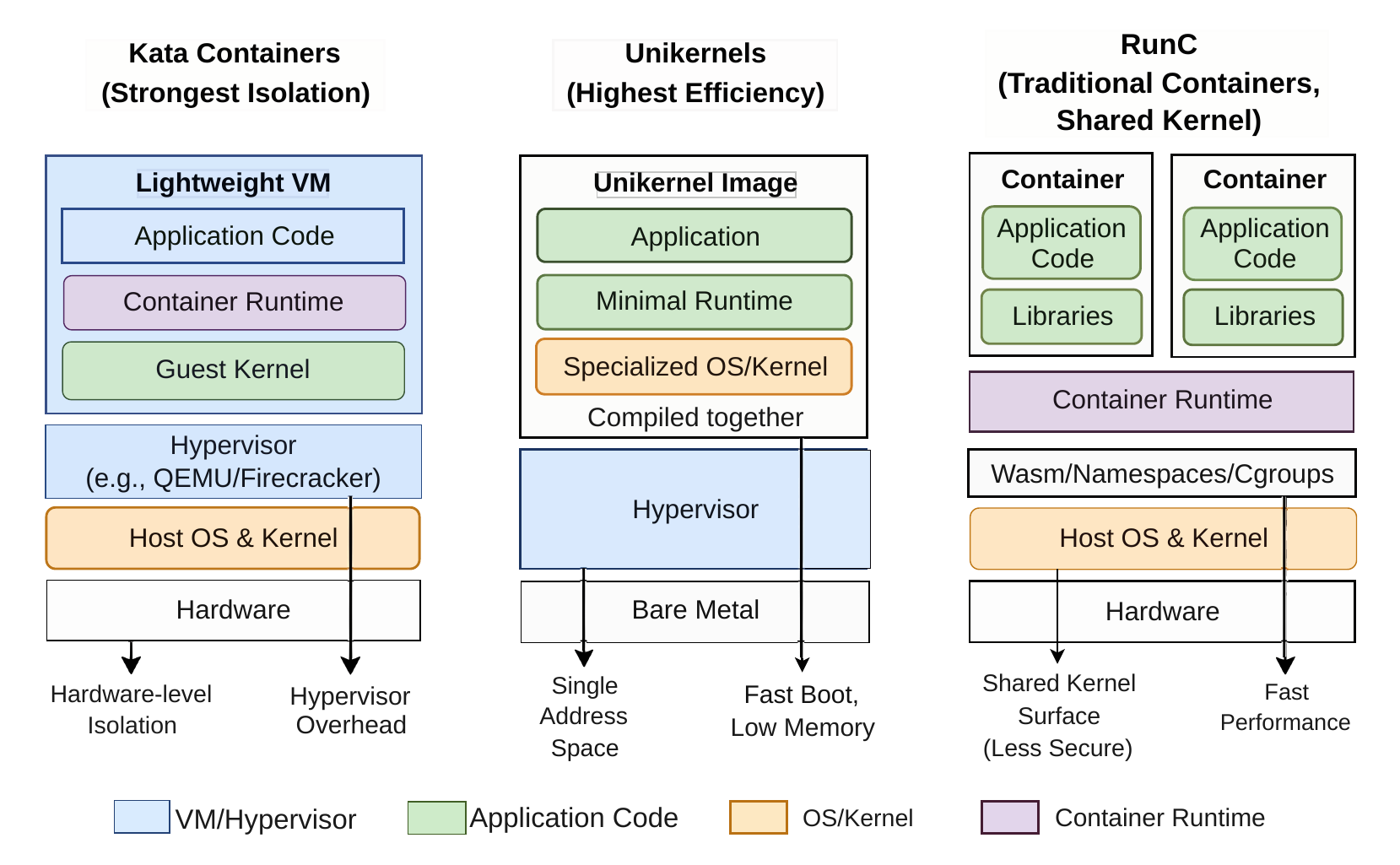}
    \caption{Comparison of different container architectures}
    \label{fig:compare}
\end{figure}

Confidential container technology is constantly evolving, aiming to provide robust isolation for sensitive workloads while maintaining performance and resource efficiency. Existing methods can be broadly categorized into four main directions, as shown in Figure \ref{fig:compare}.

\textbf{1) VM-based Isolation}: The first type is the encapsulation solution based on virtual machines or MicroVM . Represented by Kata Containers and Confidential Containers ( CoCo ), this approach runs containers inside lightweight virtual machines and leverages hardware-secured computing capabilities such as SEV, TDX , or CCA to enhance isolation. This route offers good OCI /Kubernetes compatibility and is currently the most widely adopted confidential container implementation in the industry. However, microVM encapsulation essentially continues the "containers running on top of virtual machines" structure. Due to the virtualization layer and redundant software stack, it incurs significant performance overhead , resulting in larger startup latency, significant kernel and file system redundancy, and higher resource consumption. Recent work, Crossing Shifted Moats, further points out that directly migrating existing VM -oriented confidential container designs to new hardware trust models may lead to threat model misalignment issues.

\textbf{2) Enclave Expansion}: The second category is extended solutions based on enclave/ libOS . These works , represented by SCONE, Graphene-SGX, and Occlum , primarily rely on TEEs such as Intel SGX. They deploy applications and their runtime environments within the enclave and reduce trust in the host operating system through library OS, shielding runtime, or asynchronous system call mechanisms. By minimizing the trusted computing foundation and tightly coupling resources at the application layer, the security zone extension reduces the attack surface while maintaining kernel sharing. These solutions offer lower TCBs and stronger confidentiality guarantees, but are typically limited by enclave memory size, system call complexity, and ecosystem compatibility. They are more suitable for protecting monolithic applications or specific services and are less suitable for directly supporting complex container orchestration and multi-tenant deployments in cloud-native environments.

\textbf{3) Kernel Packaging}: The third approach uses the Unikernel method. Inspired by the monolithic kernel paradigm, this method packages an application-specific kernel with containerized workloads. While it achieves strong isolation and simplifies deployment by reducing dependence on a general-purpose operating system, kernel packaging often leads to code redundancy, higher memory consumption, and reduced flexibility in managing dynamic workloads.

\textbf{4) High-level monitoring}:The fourth category is fine-grained isolation solutions for ARM TrustZone/ CCA . Represented by TZ-Container\cite{hua2021tz}, SHELTER\cite{zhang2023shelter}, and RContainer\cite{zhou2025rcontainer}, this type of research attempts to leverage security extensions on the ARM platform to push isolation granularity down from traditional virtual machines to the container level, process level, and even user-space execution level. Compared to microVM solutions, this approach better aligns with the design goals of containers—"lightweight, fast, and elastically scalable"—and better leverages the advantages of ARM CCA in hardware isolation and confidential memory. However, because it utilizes high-privilege software to enhance container security, it introduces security risks if the privileged software is compromised.

Collectively, these techniques illustrate the trade-offs between security, performance, and deployment flexibility in confidential container technologies, motivating the exploration of novel architectures that can achieve high isolation with minimal overhead.

\subsection{ARM CCA}

ARM CCA is the latest architectural extension to the ARMv9 -A specification, designed to provide fine-grained, hardware-enforced isolation for execution environments called "Realms." Unlike traditional virtualization, where guest virtual machines rely on potentially untrusted hypervisors for memory mapping and execution control, CCA divides the physical address space into separate Physical Address Spaces (PAS): secure space, insecure space, root space, and realm space. Through hardware access control and memory encryption keys for each realm space, the realm space is isolated from all software outside the realm space, including the host operating system and hypervisor.

CCA introduced the RMM, a small privileged firmware monitor running at EL2, which manages Realm lifecycle operations including memory allocation, mapping, and authentication. Realm memory is protected using a memory cryptography context, with each MEC identified by a unique memory cryptography context ID, ensuring that the contents remain cryptographically isolated even if different Realms reuse the same physical page. CCA also provides hardware-supported remote authentication, enabling tenants to verify the authenticity and integrity of the Realm execution environment before deploying sensitive workloads. By decoupling Realm isolation from untrusted hosts and aligning its threat model with confidential computing requirements, CCA creates new opportunities for designing confidential containers, avoiding the inefficiencies of VM-based encapsulation while providing robust, hardware-based security guarantees.

\begin{figure}
    \centering
    \includegraphics[width=0.7\linewidth]{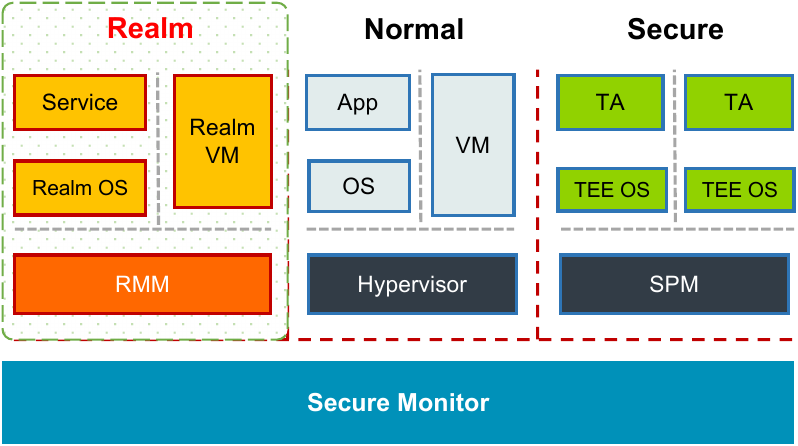}
    \caption{ARM CCA architecture}
    \label{fig:placeholder2}
\end{figure}

\subsection{Problem Analysis and Insights}

\textbf{Insight 1: Execute Containers in Realm World.} Current confidential container systems primarily execute in the Normal World, leveraging EL3 firmware for security. We believe that executing containers entirely within the Realm World (a hardware-isolated environment introduced by ARM CCA) provides a stronger protection perimeter. Realm-based execution inherits a hardware root of trust that is independent of the integrity of the Normal World software components. Once a container is instantiated within the container Realm, its execution state, memory, and page tables are cryptographically isolated from the host operating system, hypervisor, and other tenants. This shifts the TCB from a large and variable software stack to a minimal, verifiable firmware layer, aligning the system's security with confidential computing threat models.

\textbf{Insight 2: Combining High Performance with Security. } The goal is to minimize container TCB, which means running only containers within isolated Realms by directly instantiating each container as an independent Container Realm (a trend evolving from container-in-VM to container-as-isolated-domain). Containers inherently rely on a shared kernel, while Realms require independent trusted boundaries. System Realms provide shared management and service intermediaries, thus preserving the lightweight nature and cloud-native compatibility of containers while maintaining hardware-level confidentiality and integrity. Strict Realm resource partitioning ensures strong tenant isolation but makes sharing public resources (such as file caches, network stacks, or I/O buffers) difficult without compromising security. Sharing these resources across Realms can improve efficiency but also carries the risk of expanding the attack surface through shared code or data.

\subsection{Threat Model}
\label{threat_model}

We assume the existence of a powerful attacker with complete control over the host operating system, hypervisor, and all privileged software outside the domain. The attacker can execute arbitrary code at the EL1 / EL2 level, manipulate the host kernel subsystem, intercept and modify inter-virtual machine communication, and perform memory probing or DMA attacks on unencrypted memory regions . The attacker can also deploy malicious containers or virtual machines on the same physical host to launch cross-tenant attacks, including attempts to exploit shared kernel resources.
However, we trust the hardware implementation of ARM CCA, including the RMM, the CPU's memory encryption mechanisms, and the hardware-supported authentication process. We also assume that the tenant's workloads and container images themselves are not malicious, meaning they comply with established security policies and do not actively attempt to steal confidential information through covert channels . Physical attacks on the CPU package and side-channel attacks exploiting microarchitectural leaks (such as cache timings and speculative execution) are beyond the scope of this discussion , but we will discuss possible mitigation measures.

\subsection{Design Goals}
Our goal is to achieve highly secure and efficient confidential containers in cloud-native environments. Specifically, Fasco's design goals are as follows:

\textbf{G1: Code and Data Integrity. } The code, runtime state, and private data of a confidential container must remain intact throughout its entire lifecycle and cannot be tampered with by the host operating system, virtual machine monitor, or other containers. To achieve this, on the one hand, it is necessary to rely on the hardware isolation mechanism provided by ARM CCA to protect the container's execution context and memory during runtime; on the other hand, it is necessary to perform trusted loading and remote authentication of the image and runtime configuration before the container starts, so that tenants can verify that the actual runtime environment is consistent with expectations.

\textbf{G2: Data Confidentiality. } Data processed by a confidential container should remain confidential even in the presence of an untrusted host software stack. Achieving this requires: (i) securely injecting the image, keys, and input data into the container; (ii) achieving spatial isolation during execution by protecting memory and separate address spaces through Realm; and (iii) ensuring temporal isolation after the container exits to prevent sensitive data from being leaked due to page reuse, stale mappings, or improper resource reclamation.

\textbf{G3: Secure and controlled access to system services. } While confidential containers still need to access system services such as the file system and network during operation, they cannot directly trust the host OS. Therefore, Fasco needs to provide controlled system calls and I/O mediation mechanisms. Specifically, requests issued by the container Realm should be forwarded through the System Realm and a shared buffer controlled by the RMM, ensuring that cross-domain communication can only occur through explicitly authorized channels and preventing the Normal World from directly accessing the container's private memory.

\textbf{G4: Hardware-enforced resource isolation. } Even if the host OS and hypervisor are completely compromised, Fasco still needs to ensure that the memory and execution context of confidential containers are not directly corrupted.

\textbf{G5: Minimum Trusted Computing Base and Compatibility. } To reduce the attack surface, Fasco should minimize the TCB as much as possible, while reusing existing container images, OCI toolchains, and host-side infrastructure as much as possible, and avoiding large-scale modifications to the existing software stack.

\textbf{G6: Lightweight execution and fast startup with low performance overhead. } Instead of introducing a complete guest kernel and redundant virtual machine software stack for each container, maintaining the container's goal of being "lightweight and fast," it directly instantiates workloads as Container Realms, enabling them to support typical cloud-native workloads.

\section{Architecture Design}
\label{3}

\subsection{Overview}

In ARM CCA’s original design, the Realm Management Monitor (RMM) functions analogously to a virtual machine monitor (VMM) in traditional virtualization: it is responsible for creating and launching confidential virtual machines (Realm VMs), managing their memory mappings, and enforcing hardware isolation. In our design, we repurpose this capability to operate at container granularity. Instead of provisioning an entire guest OS image inside a Realm VM, the RMM directly instantiates container-level Realms, enabling finer-grained isolation with significantly lower boot and resource overheads.

Fasco’s execution model begins with the RMM creating a System Realm during system initialization. The System Realm is a privileged, yet hardware-isolated, execution environment responsible for orchestrating container lifecycle management. When a user launches a new container, the request is directed to the System Realm, which allocates CPU and memory resources, provisions a new Container Realm, and loads the container image into it. This approach decouples container instantiation from the untrusted host OS, ensuring that all critical provisioning logic runs in a trusted CCA environment.

While Container Realms are designed to run application code in complete isolation from the Normal World, many workloads require access to shared, complex, or privileged services—such as filesystem operations, network stacks, or inter-process communication—that cannot be efficiently implemented within each Realm. To address this, Fasco routes such requests through the System Realm, which delegates them to an untrusted “Trim OS” running in the Normal World. This cross-domain model allows the System Realm to leverage the Trim OS for high-performance handling of common services, while the RMM enforces strict mediation: all data and control flows are validated before re-entering the Container Realm. This ensures that untrusted components can be used to improve performance without weakening the security guarantees of Realm-based execution.

\begin{figure}
    \centering
    \includegraphics[width=0.9\linewidth]{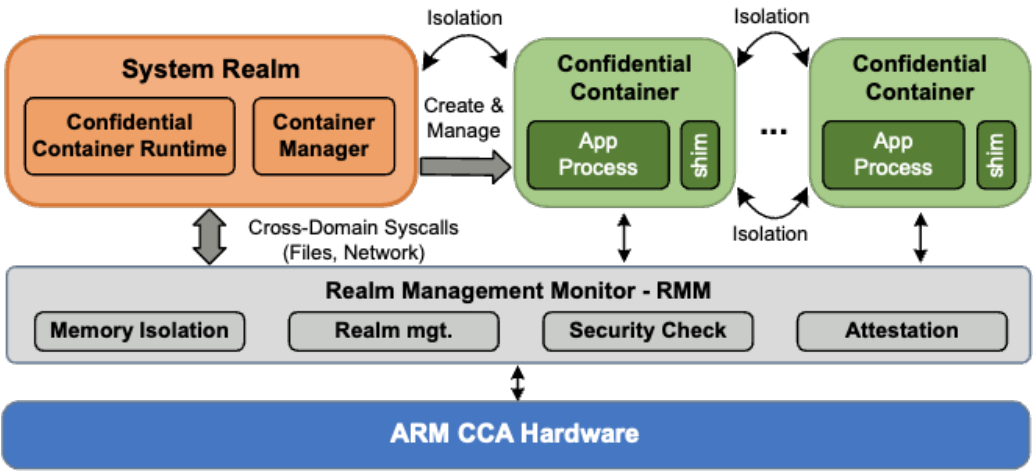}
    \caption{Design of Fasco}
    \label{fig:placeholder}
\end{figure}

\subsection{System Realm}
Container Realms require system service support to operate. One straightforward approach is to have the Container Realm directly rely on the normal operating system for system calls, file access, and exception handling. However, this design presents two problems. First, each system interaction within the container requires frequent switching between the Realm World and the untrusted Normal World, leading to significant overhead in context saving and restoration, state checks, and cross-domain authentication. Second, if the container's service requests are directly exposed to the host OS, the security of the container runtime will overly rely on untrusted software, which is inconsistent with the ARM CCA threat model.

Therefore, Fasco no longer uses the host OS as the primary service provider. Instead, it places a lightweight management and service layer, the System Realm, within the Realm World. The System Realm has two main functions: First, it receives creation requests from users, schedules the RMM through a security broker, and identifies and verifies the security of container images. Second, when faced with service requests such as system calls during container runtime, the System Realm first catches any exceptions that arise, processes them, and after the results are checked by the RMM, securely returns the data to the confidential container.

Another advantage of this design is that it allows the RMM to remain small and critical, handling only the underlying security mechanisms; while richer, yet still trustworthy, runtime logic is implemented in a separate Realm. In Fasco, the RMM is only responsible for low-level operations directly related to security, such as Realm lifecycle management, stage-2 page table isolation, context switching, exception handling, and controlled Realm entry and exit; while the System Realm handles higher-level container runtime management and service processing. This avoids the RMM from bloating into a large, difficult-to-verify, trusted kernel, and also prevents the container's critical control paths from depending on the host OS.

\subsection{Resource Isolation}
In Fasco, memory isolation is achieved by leveraging the ARM CCA Realm primitives and the Realm Management Monitor. Each confidential container is assigned a dedicated Realm, whose memory is protected using a separate set of Stage-2 page tables. Specifically, when a container Realm is created, the RMM configures a new Stage-2 translation context that maps only pages belonging to that container. These mappings come with Realm-specific protection properties, ensuring that no other Realm or untrusted system Realm can access the same physical pages. Specifically, memory isolation is considered from the following two aspects:

\textbf{1) Inter-container isolation: } Once a Container Realm's memory pages are delegated to the Realm World, they are no longer within the host's visible address space. The host OS, hypervisor, and other Normal World components cannot directly read or write these pages. 2) Normal/Secure World isolation: By reusing GPT table support in trusted firmware, an isolation boundary between the normal and secure worlds can be enforced. This ensures that memory allocated to the container domain is inaccessible to both the normal world kernel and the secure world operating system, thus maintaining confidentiality even if the host operating system is completely compromised.

\textbf{2) System Realm isolation: } Each confidential container has its own independent stage-2 translation context. A container's stage-2 page table only maps the code, data, stack, heap, and shared buffers allocated to it; the private pages of other containers remain unmapped to it. Therefore, although multiple containers reside in the Realm World, they do not share the guest kernel, address space, or stage-2 mapping domain.

\textbf{3) Normal / Secure World isolation: }
The isolation boundary between the Normal World and Secure World is enforced by reusing the Guarded Page Table (GPT) support in the trusted firmware. This ensures that memory assigned to a container realm is inaccessible to both the Normal World kernel and the Secure World OS, thereby preserving confidentiality even in the presence of a fully compromised host OS.

Specifically, Fasco allocates independent Realm granules to each container at creation time and constructs a stage-2 mapping by the RMM: only the image, runtime state, and shared buffers actually needed by the container are mapped, while the remaining pages remain unmapped. Furthermore, Fasco guarantees that a physical granule can only belong to the Normal World, the System Realm, or a single Container Realm at any given time, unless explicitly converted to a shared buffer. Container confidential pages are not aliased to the host, other containers, or the System Realm. All cross-domain interactions must be completed through the shared buffer.

\begin{figure}
    \centering
    \includegraphics[width=\linewidth]{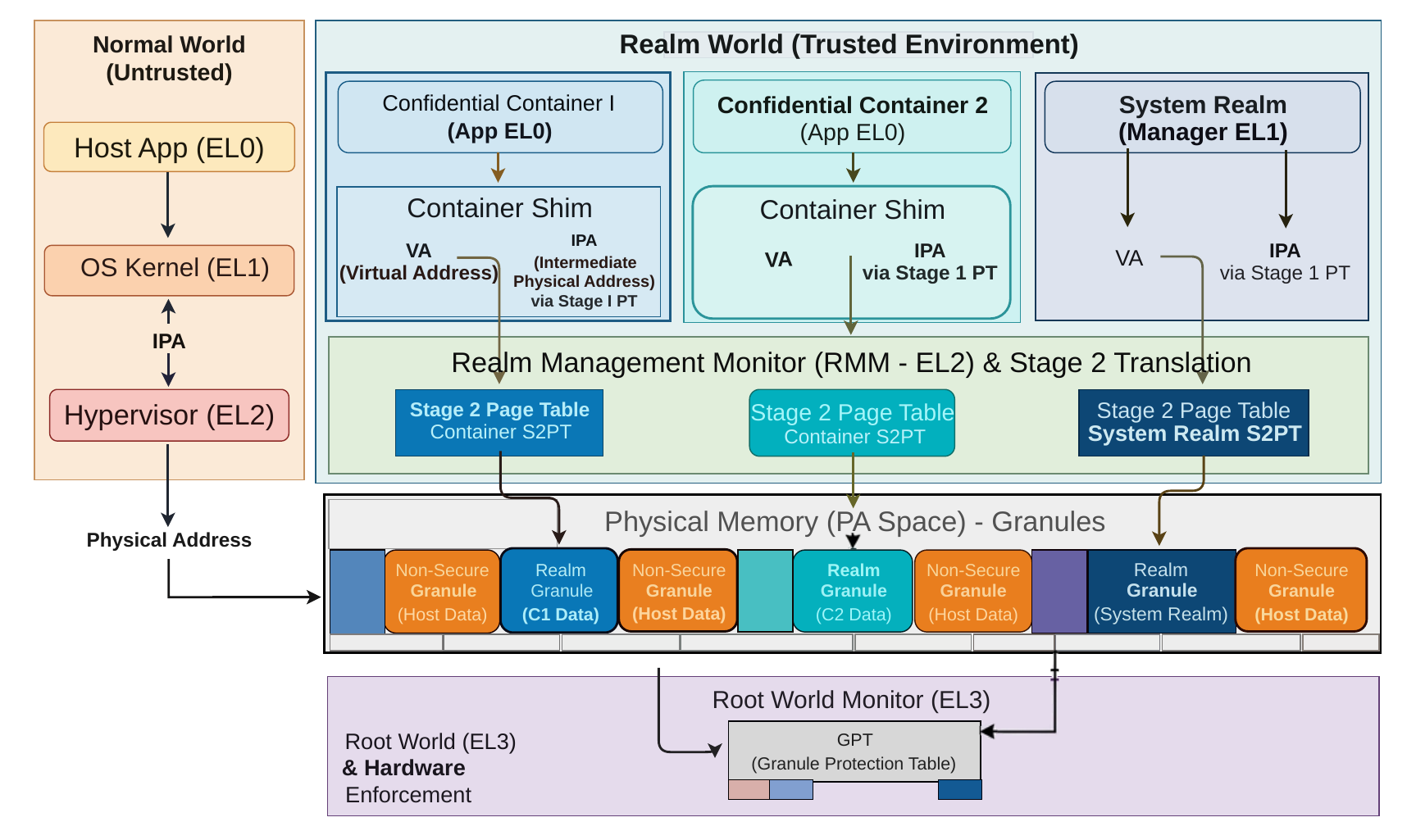}
    \caption{Container stage-2 memory isolation}
    \label{fig:memory}
\end{figure}

\subsection{Exception and Interrupt Handling}
During execution within the container domain, all interrupts and exceptions are intercepted by the realm management monitor. Upon receiving such events, the RMM securely saves the container domain's processor context, enables virtualization mode, and forwards the interrupt or exception to the system domain for handling. This design ensures that untrusted software in the ordinary world cannot directly inject into or tamper with the execution state of the container domain.

\begin{figure}
    \centering
    \includegraphics[width=\linewidth]{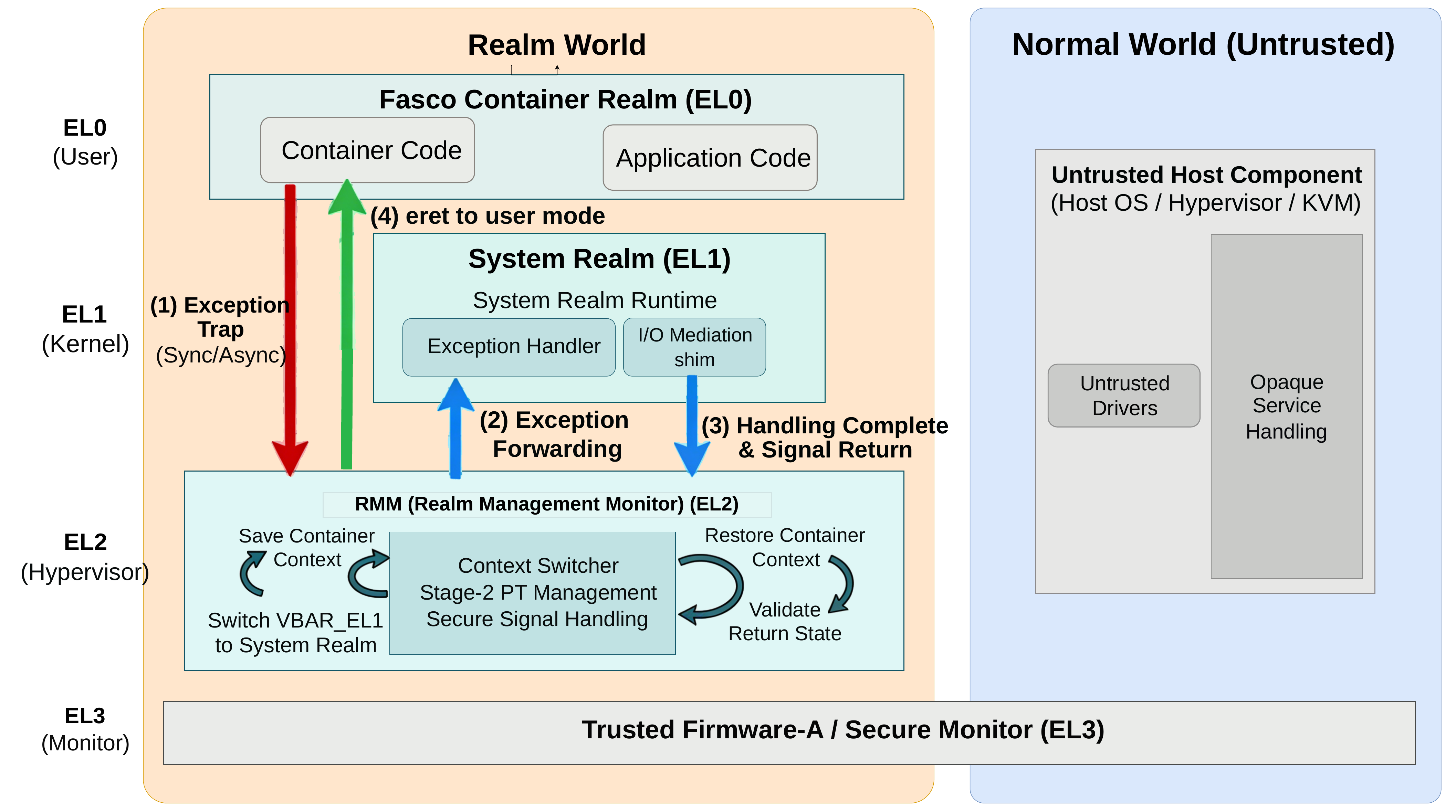}
    \caption{Exception forwarding and Interrupt handling}
    \label{fig:forward}
\end{figure}

After the system domain completes its processing routine (whether handling hardware interrupts, faults, or system calls), it invokes the \textit{eret} instruction. This instruction call enters the RMM, which restores the previously saved container context and resumes execution from the exact point where the interrupt occurred. This collaborative mechanism ensures transparent exception propagation without violating the isolation guarantees of the container domain.

\subsection{File, I/O, and Inter-Realm Communication}
Currently, many popular services, such as Redis and Memcached, are built on the premise of a trusted underlying operating system. Therefore, these services store files in plaintext, communicate with other processes via unencrypted TCP channels (i.e., without TLS), and output directly to standard output (stdout) and standard error output (stderr). To ensure confidentiality while enabling common services such as file access, network I/O, and inter-process communication (IPC), Fasco introduces a lightweight shim layer between the container domain and the system domain operating system. All I/O requests from the container domain are intercepted by this shim layer and securely forwarded to the system domain for processing. The shim's main functions include two aspects : (1) preventing underlying Iago attacks, such as the operating system kernel controlling the pointers and buffer sizes passed to services; and (2) ensuring the confidentiality and integrity of data transmitted through the operating system. Fasco supports the following protection mechanisms: (1) transparent file encryption; (2) transparent encryption of communication channels via TLS; and (3) transparent encryption of console streams. The data payload is encrypted and decrypted within a trusted boundary to prevent cross-domain leakage or leakage into the untrusted normal world.

\subsection{Cache Coherence}
The execution of confidential containers involves multiple mutually isolated execution domains: the host software stack in the Normal World, the System Realm in the Realm World, and the Container Realm that hosts the confidential container. Although these domains are strictly isolated through stage-2 page tables and address-space delegation, operations such as cross-domain shared-buffer exchange, image loading, runtime parameter copying, and page-permission updates inevitably introduce a risk of inconsistency between cache state and memory contents. Therefore, during confidential container execution, Fasco must guarantee not only memory isolation but also cache coherence during cross-domain data exchange, so as to prevent the processor from using stale cache lines, TLB entries, or instruction cache contents, which could lead to incorrect execution and even weaken isolation guarantees.

To maintain cache coherence, Fasco requires the RMM to perform the appropriate cache/TLB maintenance before control is transferred whenever cross-domain memory writes, code updates, page-table changes, or shared-data exchanges occur. Specifically, this includes three types of operations: (i) Cleaning the data cache after data-page writes, so that updated contents become visible to the target domain. (ii) Invalidating the instruction cache after code-page modifications, to prevent execution of stale instructions. (iii) Invalidating the TLB after page-table or mapping changes, so that new access permissions and address translations take effect immediately.

\section{Implementation}
\label{5}

\subsection{Runtime System}

We implement Fasco on top of ARM’s Trusted Firmware-A (TF-A) and the reference Realm Management Monitor. Our implementation introduces a small set of C/assembly modules that extend RMM to support container-level Realms and a minimal System Realm runtime. The added code covers (i) Realm-wide memory isolation, (ii) memory management (granular lifecycle, cryptographic image loading), (iii) domain switching and exception mediation (RMM trap handlers, recovery paths), (iv) a System Realm runtime (container launcher, syscall interception, policy), and (v) a streamlined I/O shim for marshalling requests to an untrusted Realm OS and performing end-to-end encryption. All trusted changes are limited to the Realm world (RMM + System Realm); Realm OS agents are untrusted and excluded from the TCB .

\subsection{RMM Modification}

To support container-level Realm execution, we have extended the existing management modules of RMM, adding RSI calls related to confidential container execution. The new RSI service interfaces include: container Realm instantiation, secure allocation of CPU and memory resources, creation of independent two-phase page tables; image decryption and verification, in-situ decryption of encrypted container images, and integrity verification before execution.

\textbf{The container creation path. }The creation handler initializes the metadata associated with the newly created confidential container. It records the container's protected memory regions, vector table locations, stack pointers, entry point, and the operating system page table base address used before entering the isolation context. Furthermore, it initializes the page table management structure for each container and allocates space for copies of the page tables used in the container's specific address space. The implementation then copies the relevant page table ranges from the host-side execution context into the container's page table structure. If the copy is successful, the new page table root is recorded as the confidential container's private TTBR0\_EL1, which will be mounted during execution switchover to the container. Therefore, this creation path establishes the execution metadata required for the isolated container to run, enabling it to use its own translation context and exception vector table.

Once the confidential container execution environment is established, but before handing control back to user space, the RMM management module performs a secure initialization sequence to prepare the container's runtime state. First, RMM backs up the operating system's exception vector table and any registers in the system domain that might be affected during container execution. This preserves the system domain's execution context and ensures that exception handling can be recovered without loss after the container's lifecycle ends. Next, RMM rewrites the container domain's Phase 1 translation table base register (TTBR0\_EL1), inserting memory mappings specific to the container execution environment. This ensures that the container can only access its own allocated memory and not map to the system domain or other containers' address spaces. Finally, RMM updates the exception vector base register (VBAR\_EL1) to point to the container domain's dedicated exception vector table. This redirection mechanism ensures that all synchronous and asynchronous exceptions are handled within the container's trusted execution context, preventing leaks or interference from an untrusted world.

\begin{figure}
    \centering
    \includegraphics[width=\linewidth]{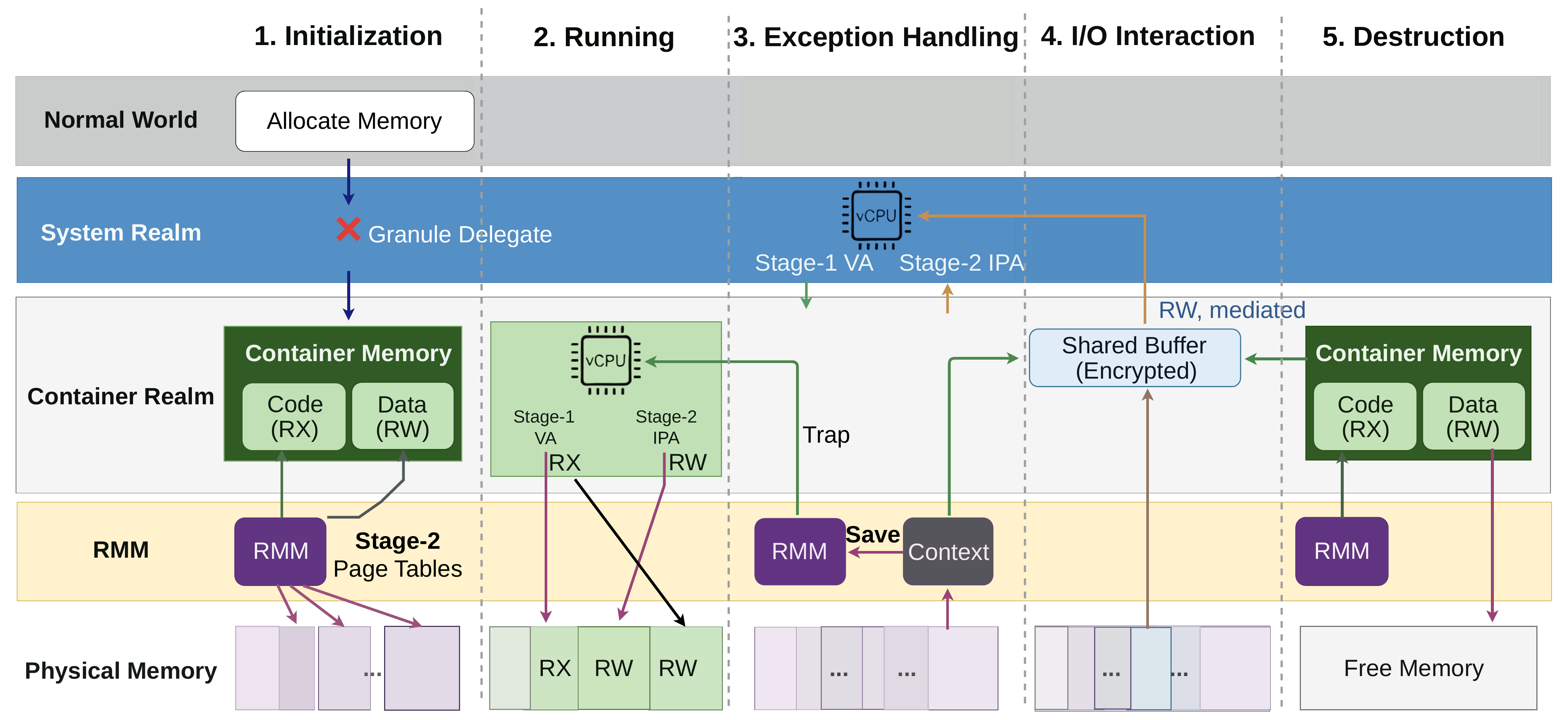}
    \caption{Memory permission change of Fasco}
    \label{fig:exception}
\end{figure}

\textbf{Two-stage page table construction.} A key requirement of Fasco is to provide an independent memory view for each confidential container without building a complete guest kernel for each container. To achieve this, the implementation copies the Realm Transition Tree ( RTT ) and derives a new second-stage transition context for each container. The cloning routine recursively traverses the source RTT hierarchy and allocates new memory granularity for the target tree from the delegated memory pool. Non-table entries are copied directly, while table entries are recreated by allocating new RTT nodes and cloning the corresponding page tables.

This design generates a structurally independent second-stage page table tree for the container. Once the cloned RTT is created, the original translation tree is modified to revoke access to the container's private memory range. Specifically, the implementation traverses the selected IPA range page by page, converting mapped allocated memory entries to allocated empty entries. Therefore, subsequent accesses to this range from the original Realm context trigger a second-stage abort, while the cloned translation tree continues to retain access for the isolated container context. This mechanism forms the basis for memory isolation between the system Realm and the container-specific execution context.

Each container Realm independently maintains its Stage-1 and Stage-2 page tables. Containers do not share any page table entries, ensuring strict isolation of address spaces. The System Realm maintains its own Stage-2 page table for the operating system and active containers. When a confidential container is created, the RMM allocates a completely new second-level page table for the container domain, which maps only the container's own memory. Simultaneously, the container's page table entries in the system domain are revoked, removing read/write permissions. Therefore, the operating system in the system domain cannot directly access the container's memory, but the container can freely access its own address space. During initialization, there is only one System Realm second-stage page table. When a confidential container is created, the RMM creates a new container second-stage page table and sets the isolated memory regions in the System Realm second-stage page table to be non-readable, non-writable, and non-executable. When the confidential container is loaded and running, it switches to this container page table, ensuring that the System Realm cannot access the container's memory. When an exception such as a system call occurs, it switches back to the original second-stage page table for execution.

\textbf{Exception and System Call Forwarding.} When a container triggers a synchronization exception, the exception handler is invoked. The handler first reads the exception syndrome register ( ESR\_EL1 ) and the fault address register ( FAR\_EL1 ) to determine the cause of the trap. The current implementation distinguishes at least two important cases. First, if the exception corresponds to a system call trap, the handler records the system call number, marks the task as waiting for the system call to return, and rewrites the system call parameters to redirect any pointer arguments pointing to container memory to an approved shared buffer. Second, if the exception corresponds to a data abort event, the handler records the fault syndrome and fault address to handle the event during the next controlled reentrancy.

The reentrant path then restores the container's exception vector base address, saves the current operating system stack pointer, mounts the container's private TTBR0\_EL1, and sets the return program counter to the instruction pointer saved by the container. Therefore, the program resumes execution in the isolated container context, with the correct address space and exception vector table. This transition is crucial: it ensures that the container always resumes execution within its own transition and exception handling environment, rather than under the control of the operating system.

\textbf{Shared Buffer Registration.} To support interaction with external services, Fasco allows containers to explicitly register communication buffers through shared handlers. The current implementation supports at least two types of shared memory: general shared data regions and signal stack regions. For each region, the handler resolves the corresponding physical address based on the provided virtual address and records the virtual address, physical address, and size in the task metadata. These shared regions are then used as the basis for communication of mediated system calls and asynchronous events. By requiring explicit registration, Fasco avoids exposing arbitrary container-private pages to the external operating system. Instead, only specified buffers can perform controlled cross-domain data exchange.

\textbf{Cache Coherence.} In Fasco, cache maintenance should be placed within the trusted execution chain formed by the RMM and the System Realm, corresponding to three specific scenarios. First, container image loading and runtime initialization. After the System Realm writes image pages, metadata, or entry code into the private pages of a Container Realm, it should first perform a D-cache clean. If these pages are going to be executed, it must further perform an I-cache invalidate, and, when necessary, use appropriate barriers to ensure that the container observes the most up-to-date code when execution resumes. Second, shared-buffer communication. When the Container Realm and the System Realm exchange requests and return results through shared pages, the writer must complete cache cleaning before the context switch so that the receiver can observe a consistent view of the data. If the host participates in controlled I/O sharing, it may do so only through explicitly shared pages and must follow the same synchronization rules. Third, stage-2 page table and sharing-attribute updates. Once the RMM or the System Realm establishes, modifies, or revokes a stage-2 mapping for a Container Realm, it must invalidate the relevant TLB entries accordingly, to prevent the container or any other execution path from continuing to use stale permissions or outdated translation results.

\section{Security Analysis}
We analyze Fasco under the threat model in Section \ref{threat_model}, focusing on confidentiality and integrity in the presence of a compromised host OS and hypervisor.

\textbf{Confidentiality and integrity of container execution.}
Fasco runs each confidential container inside a dedicated Container Realm, with its own stage-2 translation context and Realm-protected memory managed by the RMM. As a result, the host OS, hypervisor and other containers cannot directly access plaintext container memory. Fasco also revokes the System Realm’s access to container memory after initialization. In addition, control-state transitions are mediated by the RMM: page tables, execution context, and exception vectors are initialized and protected within the Realm world, and exceptions are trapped and forwarded through trusted paths. This prevents a compromised Normal World from tampering with container memory or execution state.

\textbf{Inter-container isolation.}
Each container is assigned an independent Realm with separate stage-2 page tables and disjoint encrypted memory regions. No private plaintext memory or page tables are shared across Container Realms. Therefore, even if one container is compromised, it cannot directly access the memory or execution state of another container. This design removes the shared-kernel attack surface of conventional containers and strengthens cross-tenant isolation.

\textbf{Protection against an untrusted host.}
Fasco places security-critical operations, including container creation, protected memory assignment, exception mediation, and attestation support, inside the Realm world. Thus, even a fully compromised host kernel or hypervisor cannot directly violate the confidentiality or integrity of workloads inside Container Realms. At most, the host can cause denial of service by delaying scheduling, withholding resources, or dropping requests.

\textbf{Secure mediation of system services.}
Since Container Realms cannot trust the host OS for direct system services, Fasco mediates file, network, and other I/O operations through the System Realm and shared buffers. Trusted components validate buffer metadata, ownership, and result sizes before data are returned to the container, reducing the risk of Iago-style and interface-manipulation attacks. In addition, Fasco protects sensitive data crossing trust boundaries through shielded I/O mechanisms such as encrypted files, channels, and console streams.

\section{Evaluation}
\label{6}

\subsection{Experimental Setup}

For the experiment setup, we conducted simulations to test the functionality and performance of confidential containers. The tests took place on two platforms: Qemu and RK3588. In the Qemu environment, we set up a simulation using the Ubuntu 22.04 operating system to replicate the ARM CCA environment. This allowed us to run the project prototype and execute scenarios with the designed confidential containers. On the RK3588 development board~\cite{bertschi2025opencca}, we focused on transferring and validating the performance gains of the confidential container architecture, comparing it against traditional solutions such as the Kata container. The results demonstrated a performance improvement while maintaining security standards, underlining the potential of this technology for real-world application in cloud-native security environments.

Our evaluation aims to answer the following questions.
\begin{enumerate}[leftmargin=*]
    \item \textbf{RQ1: Baseline Environment Benchmark.} What are the baseline performance benchmarks for the confidential container under a standard computing environment without any additional security features enabled?(\cref{sec1})
    \item \textbf{RQ2: Impact on Container Runtime Performance.} How does the implementation of confidential containers affect the runtime performance compared to traditional container technologies in terms of speed, efficiency, and resource utilization? (\cref{sec2}): 
    \item \textbf{RQ3: Impact on Container Lifecycle.} : What are the implications of confidential container technology on the lifecycle of containers, from creation to destruction, including setup time and operational stability?(\cref{sec3})
    \item \textbf{RQ4: Impact on Multi-Container Environments} How does the performance and security of the confidential containers scale within multi-container environments, especially in terms of resource allocation, isolation, and potential interference between containers? (\cref{sec4})
\end{enumerate}

\subsection{RQ1: Baseline Environment Benchmark} 
\label{sec1}

\textbf{(1) UnixBench.} Fasco achieves a System Benchmarks Index Score of 419.2, whereas Coco scores significantly lower at 173.6. This indicates that Fasco generally outperforms Coco across the measured benchmarks.

Both Fasco and Coco show strong performance in the Dhrystone and Double-Precision Whetstone tests, which measure integer and floating-point CPU performance. Fasco has slightly higher index values in these tests, implying marginally better CPU execution efficiency. In file copy operations, Fasco substantially outperforms Coco. The indices (such as for "File Copy 1024 bufsize 2000 maxblocks") are considerably higher in Fasco, indicating more efficient file input/output operations. This may suggest better optimization and resource management in file handling for Fasco.

\begin{figure}[hbp]
    \centering
    \includegraphics[width= \linewidth]{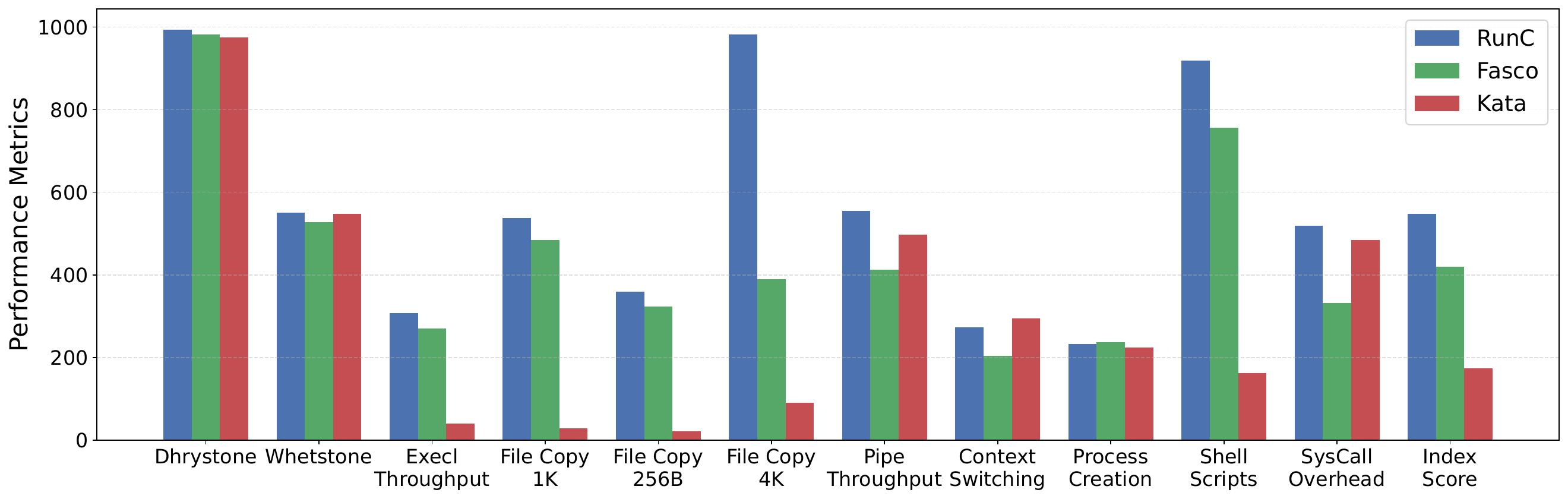}
    \caption{UnixBench Test Result}
    \label{fig:placeholder3}
\end{figure}

\begin{figure*}[htbp]
    \centering
    \includegraphics[width=\linewidth]{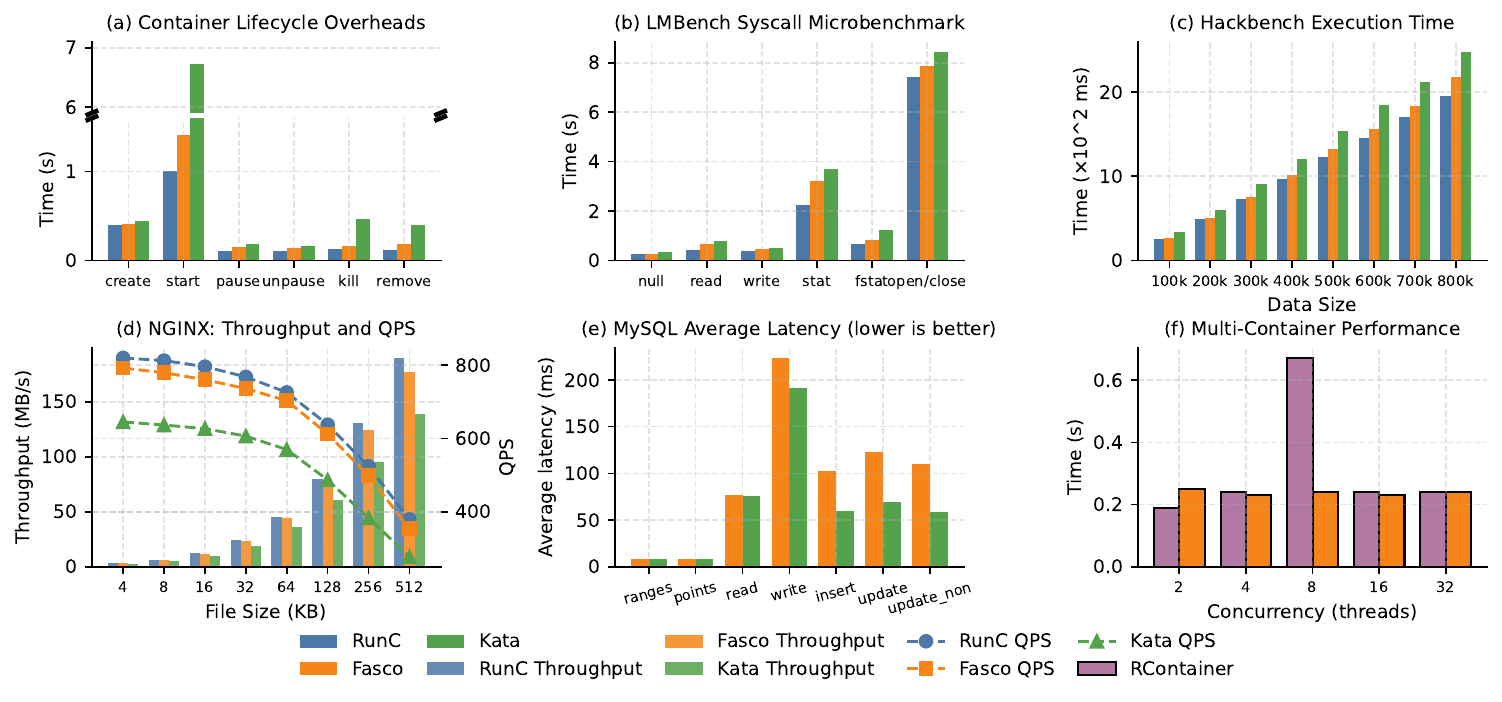}
    \caption{Micro-references performance overhead and application workloads}
    \label{fig:all}
\end{figure*}

Fasco again demonstrates superior performance in pipe throughput and pipe-based context switching, with higher index scores. These metrics are essential for evaluating multitasking and inter-process communication efficiency, areas where Fasco's architecture appears to provide enhanced throughput. Both testing environments show similar capabilities in process creation and the execution of shell scripts, although Fasco registers a better performance, reflecting its competent design for handling concurrent and sequential script processing seamlessly. Fasco excels in reducing system call overhead, suggesting lower latency and improved efficiency in handling system calls. This aspect is crucial for high-performance applications requiring frequent interactions with the OS.

\textbf{(2) LMBench} we conducted a series of benchmarks using LMBench, comparing it to other container runtimes such as RunC and Kata, alongside a baseline labeled as "Normal." The results provide key insights into Fasco's performance characteristics and its relative positioning compared to conventional container architectures. This indicates that Fasco's architecture efficiently manages its fundamental operations, successfully minimizing the additional complexity and overhead typically associated with increased security measures.

For the "read" and "write" benchmarks, which test the system's I/O performance, Fasco exhibits a moderate increase in latency over RunC and the Normal environment. This is expected, given the security layers involved in intercepting and handling system calls within the Secure Realm. Despite these additional steps, the performance remains within acceptable limits, highlighting Fasco's capability to maintain efficiency without substantial degradation in I/O-intensive operations. In the "open/close" benchmark, Fasco's results are closer to those of RunC, showing only a minor increase in latency. This indicates that the Fasco architecture can efficiently handle frequent and repeatable operations with a manageable overhead, achieving a near par performance with the traditional container runtime.

\subsection{RQ2: Impact on Container Runtime Performance} 
\label{sec2}

\textbf{(1) HackBench} Docker serves as the baseline in this performance comparison. It demonstrates relatively stable performance across different workload sizes (ranging from 100k to 800k), maintaining execution times from 1.52 to 11.08 (×100ms), a testament to its well-optimized and widely-used container runtime. While Docker is not specifically designed for confidential computing, its performance serves as a benchmark for RunC, Fasco, and Kata.

The RunC container runtime shows higher execution times, suggesting increased overhead perhaps due to isolation features which cater to improved security at the performance cost. RunC's times range from 2.59 to 19.48 (×100ms) across the same workload sizes. Compared to Docker, RunC introduces notable latency, indicating a trade-off between efficiency and security enhancements.

Fasco, the focus of this study due to its novel incorporation of the CCA features, delivers performance times from 2.62 to 21.80 (×100ms). Although similar to RunC, Fasco's architecture leverages Container Realms to execute each container separately, providing intrinsic confidentiality and integrity protections without significant performance compromise. Fasco's design balances isolation and resource efficiency, demonstrated by its competitive performance metrics. Coco, which is known for its security-focused container runtime, reports execution times from 3.42 to 24.73 (×100ms), clearly indicating the most considerable overhead among the examined technologies. This is consistent with Kata's architecture aimed at maximizing security through virtualization techniques, which introduces latency as a trade-off for achieving comprehensive protection.

\textbf{(2) Nginx} RunC generally exhibits higher throughput and QPS values across all file sizes compared to Fasco and Kata. At smaller file sizes like 4KB and 8KB, RunC achieves a QPS of approximately 820-813 with throughput maintaining around 3.35 to 6.5 MB/s. These values show RunC's ability to process a large number of small requests without significant degradation in speed, which is beneficial for applications dealing with numerous small-sized transactions.

Fasco, while slightly underperforming compared to RunC, demonstrates competitive throughput and QPS metrics. With increasing file sizes, its performance remains relatively stable, especially in scenarios where confidentiality and integrity are prioritized. At a 4KB file size, Fasco achieves a QPS of 792 and throughput is around 3.09 MB/s, close to RunC's performance. 

Kata exhibits the lowest throughput and QPS across almost all file sizes, reflecting its design focus on security isolation over raw execution performance. This trend suggests that Kata containers may incur more overhead when managing web server loads, likely due to additional security mechanisms inherent to its architecture. Despite its lower performance, Kata's robust security could be invaluable for tasks where high assurance levels are essential, outweighing the need for maximum throughput.

\subsection{RQ3: Impact on Container Lifecycle} 
\label{sec3}

In container creation times, all systems perform similarly, with Docker slightly leading with the quickest time of 0.386 seconds, closely followed by RunC at 0.402 seconds, Fasco at 0.412 seconds, and Kata at 0.446 seconds. The added latency in Fasco is minimal compared to its security advantages, indicating an efficient initialization process despite the inherent complexity of leveraging ARM's CCA for secure memory and isolation.

When examining the start times, Fasco demonstrates a moderate increase in time to 1.406s compared to RunC's 1.006s, yet it significantly outperforms Kata, which lags with a noticeable delay at 6.727. Docker remains the fastest at 0.431 seconds. Fasco's performance in this area suggests an effective integration of CCA's Realm capabilities without the substantial overhead visible in Kata, which likely results from Kata's more complex virtualization layer.

During pause and unpause operations, Fasco shows a slightly longer pause time at 0.152 seconds compared with RunC and Docker, both at 0.102 seconds. However, it matches Kata's pause time of 0.186 seconds. Unpause times for Fasco and Kata are identical at 0.139 seconds, with RunC and Docker being quicker. These results imply that while Fasco introduces some overhead, it maintains a consistently competitive edge against Kata, highlighting its efficient handling of temporary state changes within the Realm architecture.

The 'kill' operation is particularly illustrative of Fasco's optimized state management, achieving the fastest time among the secure designs at 0.162 seconds. This is notably quicker than Kata's 0.469 seconds and even surpasses Docker, which consistently outpaces others in previous categories with a time of 0.124 seconds. Here, the Fasco design shows its capability to swiftly manage and terminate container processes, likely benefiting from its direct Realm deployment strategy.

Finally, in the remove operation, Fasco demonstrates superior performance in managing container teardown, with a moderate completion time of 0.189 seconds. This is significantly faster than Kata's 0.399 seconds but slightly behind RunC and Docker, pointing to some overhead associated with securely dismantling Realm containers while maintaining isolation integrity. However, Fasco's performance here remains commendable given its dedicated focus on security through hardware-enforced methods.

\subsection{RQ4: Impact on Multi-Container Environments} 
\label{sec4}

Fasco demonstrates highly consistent performance across all tested scenarios, with recorded overhead values remaining within a very narrow range of 0.23–0.25 seconds. Unlike RContainer's spike at 8 containers, Fasco's overhead does not fluctuate with the increasing container count, which could indicate better resource management under concurrent container workloads. Similar to the original discussion, the limitations of the development board or simulator could influence these results. For example, hardware resource constraints or simulation inaccuracies (especially for RContainer, given the ARMv9-related simulation) might cause deviations in performance measurements. Particularly for RContainer, its simulated environment on ARMv8 hardware instead of real ARMv9-A hardware might introduce non-deterministic variability, potentially explaining the overhead deviation observed at 8 containers.

\section{Related work}
\label{8}

\newcommand{\posc}{\tikz\fill[black] (0,0) circle (2pt);} 
\newcommand{\negc}{\tikz\draw[black] (0,0) circle (2pt);} 
\newcommand{\na}{--}  

\begin{table*}[t]
\centering
\scriptsize
\setlength{\tabcolsep}{5pt}
\renewcommand{\arraystretch}{1.15}
\caption{Comparison of container/TEE systems.}
\begin{tabular}{ll|cccc|c|c|c}
\toprule
\multicolumn{2}{c|}{} &
\multicolumn{4}{c|}{\textbf{Security}} &
\multirow{2}{*}{\textbf{Compatibility}} &
\multirow{2}{*}{\textbf{Performance}} &
\multirow{2}{*}{\textbf{Trusted Components}} \\
\multicolumn{1}{c}{\textbf{Enclave Type}} &
\multicolumn{1}{c|}{\textbf{System}} &
MUMA & Iago & Abs.\ resource & Physical & & & \\
\midrule
\multirow{5}{*}{\textbf{Normal World}}
& RContainer\cite{zhou2025rcontainer}    & \posc & \posc & \posc & \negc & \posc & $\sim5.7\%$ & TF-A$^{+}$ (EL3) + mini-OS (EL1) \\
& Shelter\cite{zhang2023shelter}       & \negc & \posc & \negc & \negc & \posc & $<15\%$ & TF-A$^{+}$ (EL3) \\
& BlackBox\cite{van2022blackbox}      & \negc & \posc & \negc & \negc & \negc & $<15\%$ & CSM (EL2) \\
& TZ-Container\cite{hua2021tz}  & \posc & \posc & \negc & \negc & \posc & $\sim5\%$ & TF-A$^{+}$ (EL3) \\
& Sanctuary\cite{brasser2019sanctuary}     & \negc & \posc & \negc & \negc & \posc & N/A & OP-TEE (S-EL1) + TF-A$^{+}$ (EL3) \\
\midrule
\multirow{3}{*}{\textbf{Realm World}}
& Aster\cite{kuhne2024aster}         & \negc & \posc & \negc & \posc & \posc  & $\sim10\%$ & RMM$^{+}$ (R-EL2) + TF-A (EL3) \\
& Coco/Kata\cite{coco}     & \posc & \na & \na & \posc & \posc & $\sim40\%$ & OS (EL1) + RMM (R-EL2) + TF-A (EL3) \\
\cmidrule(lr){2-9} 
& \textbf{Fasco} & \posc & \posc & \posc & \posc & \posc & $\sim12\%$  & \textbf{RMM$^{+}$ (R-EL2) + TF-A (EL3)} \\
\bottomrule
\end{tabular}
\end{table*}

\textbf{Arm TEEs.} In this paper, we present Fasco, a scalable isolation architecture on Arm CCA. Therefore, we focus on discussing related TEE works on the Arm platform. vTZ ~\cite{hua2017vtz} leverages the stage-2 translation in the Normal World to implement secure virtualization.  Container security enhancements based on TEE. Haven~\cite{baumann2015shielding}, Graphene-SGX~\cite{tsai2017graphene}, Occlum~\cite{shen2020occlum}, and others typically deploy libOS within SGX containers to reduce the attack surface. SCONE~\cite{arnautov2016scone} runs containers only within the enclave and improves container performance through asynchronous system calls. Hua et al. proposed TZ-Container\cite{hua2021tz}, which uses ARM TrustZone to protect containers from operating system attacks, effectively preventing MUMA attacks. Sanctuary ~\cite{brasser2019sanctuary} uses TZASC to partition different address spaces and build user-space isolated execution environments like enclaves. BlackBox~\cite{van2022blackbox} deploys a CSM within the normal world EL2 to prevent containers from being attacked by the operating system. The Split Container~\cite{shi2025serverless} approach separates the "secure execution plane" and the "management plane," using a function-oriented OS of "microkernel + LibOS" to host multiple serverless functions within the CVM, while the management plane still reuses the untrusted Linux, achieving significant performance improvements in function scenarios.

\textbf{Confidential Containers.} Currently, the mainstream confidential container~\cite{coco} approach uses the Kata scheme, deploying confidential containers directly within virtual machines. Valdez et al. ~\cite{valdez2024crossing} points out that this approach may lead to attacks due to the misalignment of the normal world and Realm World threat models. Shelter~\cite{zhang2023shelter} utilizes the hardware provided by ARM CCA to build multiple GPT tables and run trusted applications. Modifying the code in EL3 may increase the attack surface. To address this risk of attack surface expansion introduced by EL3 modifications, Zhou et al. proposed RContainer~\cite{zhou2025rcontainer}, a mini-OS that places some trusted code in EL1, thereby reducing the attack surface. PORTAL~\cite{sang2025portal} addresses the device I/O security and efficiency challenges of mobile SoCs by proposing secure I/O through memory isolation in ARM CCA. Leveraging CCA's hardware-level access control, it ensures that only designated Realm virtual machines and peripherals can access plaintext memory areas protected by PORTAL, thus avoiding the performance and energy overhead of encryption and decryption. Currently, CCA hardware is difficult to obtain. TwinVisor ~\cite{li2021twinvisor} and virtCCA~\cite{xu2023virtcca} utilize the secure EL2 support in TrustZone to simulate CCA and run confidential virtual machines.

\section{Conclusion}
\label{9}

This paper presents \textit{Fasco}, a lightweight confidential container runtime built on ARM CCA. Unlike existing confidential container designs that encapsulate each container inside a microVM, Fasco directly instantiates each container as an independent Container Realm and introduces a System Realm to provide controlled resource management and system service mediation. By leveraging RMM-based memory isolation, exception forwarding, and shared-buffer communication, Fasco achieves hardware-enforced confidentiality and integrity protection for container workloads while avoiding the heavyweight software stack of VM-based approaches.

We implemented a prototype of Fasco and evaluated it on both a QEMU CCA environment and an ARM-based hardware platform. The results show that Fasco can reduce startup latency and runtime overhead compared with existing microVM-based confidential container solutions, while maintaining a relatively small TCB and preserving compatibility with containerized workloads.

\bibliographystyle{ACM-Reference-Format}
\bibliography{fasco}

\appendix 

\end{document}